\documentclass[aps,twocolumn,10pt,prl,nofootinbib,showpacs]{revtex4-1}
\usepackage{latexsym}
\usepackage{amsmath}
\usepackage{amsfonts}

\usepackage{hhline}
\usepackage{graphicx}
\usepackage{amssymb}

\usepackage{multirow}
\usepackage{rotating}
\usepackage{animate}

\usepackage{comment}

\usepackage{xr-hyper}

\usepackage{color}
\usepackage[normalem]{ulem}

\newcommand{\unit}[1]{\ensuremath{\, \mathrm{#1}}}

\begin{document}
\title{Small activity differences drive phase separation in active-passive polymer mixtures}
\author{Jan Smrek}
\email{smrek@mpip-mainz.mpg.de}
\author{Kurt Kremer}
\email{kremer@mpip-mainz.mpg.de}
\affiliation{Max Planck Institute for Polymer Research, Ackermannweg 10, 55128 Mainz, Germany}

\begin{abstract}
Recent theoretical studies found that mixtures of active and passive colloidal particles phase separate but only at very high activity ratio. The high value poses serious obstacles for experimental exploration of this phenomenon. Here we show using simulations that when the active and passive particles are \emph{polymers}, the critical activity ratio decreases with the polymer length. This not only facilitates the experiments but also has implications on the DNA organization in living cell nuclei. Entropy production can be used as an accurate indicator of this non-equilibrium phase transition.
\end{abstract}

\maketitle

Active matter consists of microscopic constituents, such as bacteria, micro-swimmers or molecular motors that transform energy from the surroundings or their own sources into mechanical work. The energy is fed into the system on the particle scale producing local spatio-temporal gradients which give rise to a number of interesting macroscopic non-equilibrium phenomena such as emergence of spatial structures through dynamical phase transitions \cite{Palacci_active_colloids}, directed rotational motion \cite{DiLeonardo_ratchet_motors}, propagating waves \cite{Schaller_active_filaments_NAT2010, Szabo_swarming_cells} and phase separation of mixtures of active and passive particles with purely repulsive interactions \cite{Cates_active_passive_separation, Frey, Frey_polymers, Moshing_PRL, Trefz_Binder, Joanny_Grosberg}.

The latter phenomenon, that we focus on here, is particularly interesting as often the active particles reside in a passive environment or the level of their activity is heterogeneous. Moreover, the mixture of active and passive polymers was hypothesised to play a role in the DNA organization within the cell nucleus during the interphase (metabolic phase of the cells life) \cite{Ganai_active_DNA_simulation,Awazu_active_polymer_separation,Grosberg_loop_extrusion_comment}. Indeed, DNA transcription or the hypothesised active loop extrusion \cite{Goloborodko_loop_extrusion_model} involve continual energy influx and dissipation on microscopic length scales. Observations of living cells \cite{Cremers_Active_Inactive_nuclear_organization,Solovei_chromatin_organization} and the ``C-experiments'' \cite{Lieberman_HiC} confirm that euchromatin - the active DNA, is colocalized and separated from the inactive denser heterochromatin by an unknown mechanism. The chemical difference of the hetero- and euchromatin \cite{Solovei_chromatin_organization} could play a role in the chromatin separation \cite{Jost_epigenome_folding}, and, as first shown by Ganai et al \cite{Ganai_active_DNA_simulation}, the active process that we are focusing on here too, could bring an additional contribution to the separation.

Recently, two works \cite{Frey, Joanny_Grosberg} modelled active colloidal particles as having higher diffusivity as their passive counterparts, by simply connecting them to a higher temperature thermostat. Both, simulations and analytical theory predict phase separation at a temperature ratio of about $30$.
Similarly, high critical activity contrast is observed also for other, ``vectorial'' activity models such as active Brownian particles (ABP) \cite{Cates_active_passive_separation} (P\'{e}clet number over $50$) and Vicsek model \cite{Vicsek,Trefz_Binder} where the generated force or velocity acting on the particle has a specific direction that randomises on longer time scales. These models are more complex due to other effects like pressure dependence on the interaction details of the particles with the container \cite{Cates_Kardar_pressure_active_matter_NATURE,Kardar_Kavli_talk,Pressure_wall_curvature}. 
Thus we here restrict ourselves to the simple but rich two temperature model and apply it to polymers. With the DNA application in target, the work \cite{Ganai_active_DNA_simulation} observed phase separation of partly active polymers, where some isolated monomers had temperature $20$ times higher than the ambient one.  Although hydrolysis of a single molecule of ATP releases energy of about $20 k_{B}T$, this can not be completely spent on the increase of the kinetic energy of the DNA. In any case, the factor of $20$ or $30$ of the critical activity contrast means, the active component in room temperature environment would have a temperature of about $10^{4}\unit{K}$ - certainly beyond biologically relevant scale of intracellular processes.

However, the polymer degrees of freedom of the particles can play a crucial role in decreasing the critical activity ratio to relevant scales. It is well known that in \emph{equilibrium} phase separation of \emph{polymer} mixtures, the critical Flory interaction parameter, characterising the asymmetry of the interaction between the two polymer species and driving their segregation, is inversely proportional to the polymer length $N$ \cite{Rubinstein_book}. There, to leading order, the entropy, driving the mixing, is proportional to number of chains $M$, while the interaction, favouring the separation is proportional to $MN$, resulting in the $1/N$ dependence of the critical interaction parameter found in Flory--Huggins theory. Naturally one asks: Is there an analogous effect in the active polymer matter? In other words, are for long (bio)polymers small activity differences sufficient to drive phase separation? This is what we set out to investigate.

To do so, we simulate monodisperse polymer melts of standard fully flexible and weakly semi-flexible bead spring systems \cite{KG}, detailed in Supplementary Material (SM) \cite{supp_mat}, of $M=1000$ chains of $N$ beads that have been frequently used to investigate \emph{equilibrium} properties of dense polymeric systems. We investigate systems of $N=10$, $20$, $40$, $70$ and $100$ for flexible and additionally $N=200$ for semi-flexible chains at constant volume with bead density $\rho = 0.85\sigma^{-3}$  and periodic boundary conditions. Chains are effectively uncrossable and the entanglement length is $N_{e} \simeq 87$ in the flexible and $N_{e}\simeq 28$ in the semi-flexible system \cite{Livia_equlibration_long_melts}. To model activity differences, $M/2$ ``cold'' chains are coupled to a Langevin thermostat with temperature $T_{c} = 1.0 \epsilon$ and the other ``hot''  chains to a Langevin thermostat with $T_{h} \geq T_{c}$. Each chain is either completely hot or cold, different from \cite{Cugliandolo_active_polymers_1,Cugliandolo_active_polymers_2} where only the central monomer is active. The coupling with the thermostats is through friction coefficient $\gamma$. The choice of $\gamma$ and the time step, besides the required stability of the simulation, is connected to the fact that the energy dissipation governed by $\gamma$ is relevant for the specific interaction strength due to the heat transfer between the two species. Since we are here looking at the chain length effect only, one should use the same $\gamma$ and time step for all the temperatures and chain lengths. This is satisfied by a rather high friction coefficient $\gamma = 10 \tau^{-1}$ and a time step $\Delta t = 0.005\tau$. For comparison, the simulations in \cite{Frey_2temp_phase_separation_PRL2016} used Brownian dynamics which corresponds to a high friction limit of molecular dynamics (MD) that we do here. All our simulations were performed using the ESPResSo++ simulation package \cite{Espressopp}. As our polymers differ only by the thermostat, their phase separation and it's behaviour for different $N$ is a pure non-equilibrium effect arising from their different kinetic properties (stiffness difference due to the temperature contrast is negligible - see SM).

We start from well mixed equilibrium configurations ($T_{h} = T_{c}$), then increase $T_{h}$ and let the system evolve until a steady state is reached, typically at a time chains need to diffuse through the whole system. In the steady state:

$(i)$  For high $T_{h}$ we find two liquids (as both components are well above the glass transition point) of different hot-cold composition. For temperatures close to the onset of segregation we observe slowing down of the evolution and large fluctuations (Fig.~S1 in SM). 

\begin{figure}[htb]
\includegraphics[width=\columnwidth]{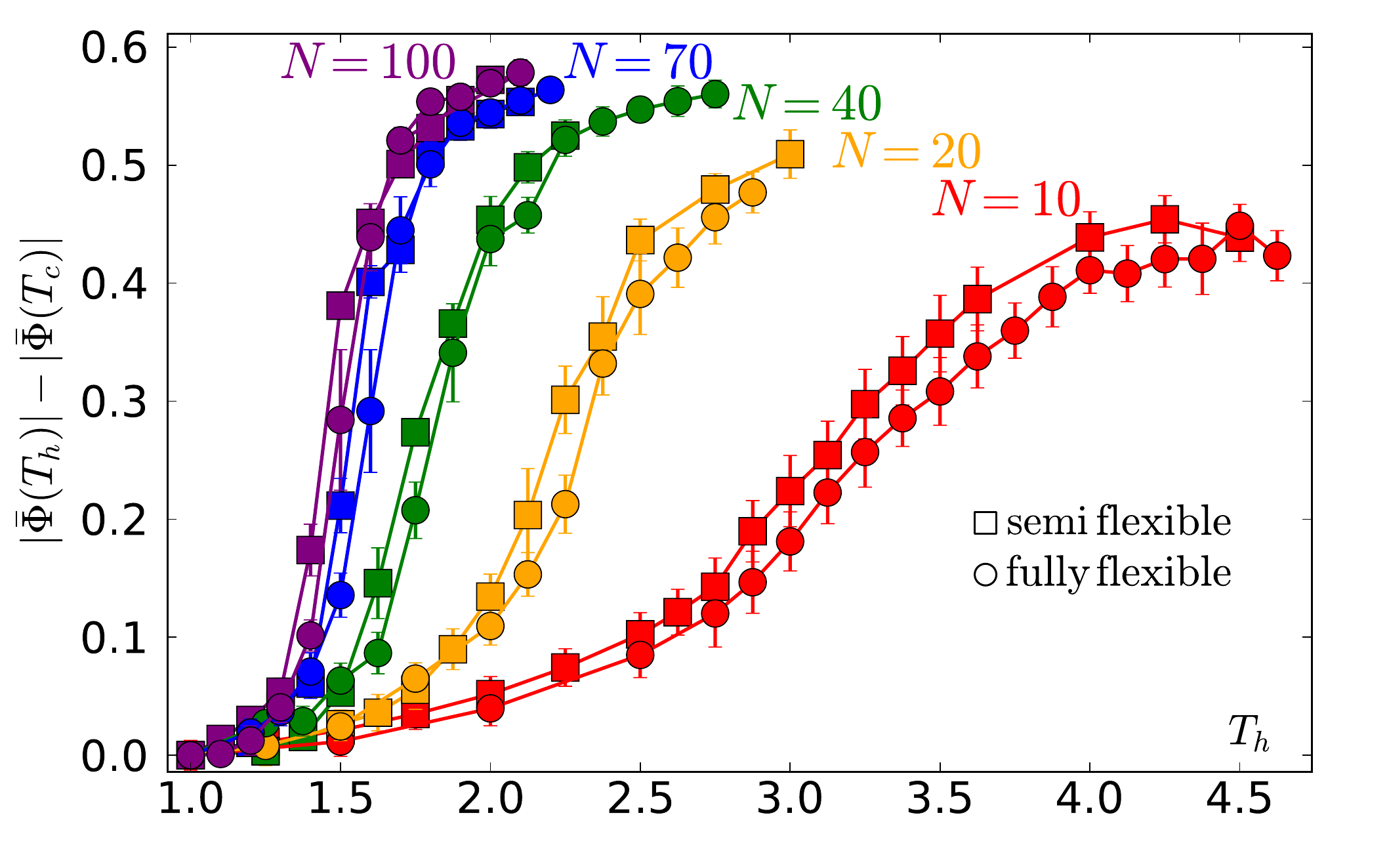}
  \caption{(Colour online) Order parameter $|\bar{\Phi}|$  as function of $T_{h}$ and $N$. A comparison of semi-flexible (squares) and fully-flexible (circles) chains shows systematically lower transition temperature for semi-flexible polymers.}
\label{fig_phi_vs_T}
\end{figure}

\begin{figure}[htb]
\includegraphics[width=\columnwidth]{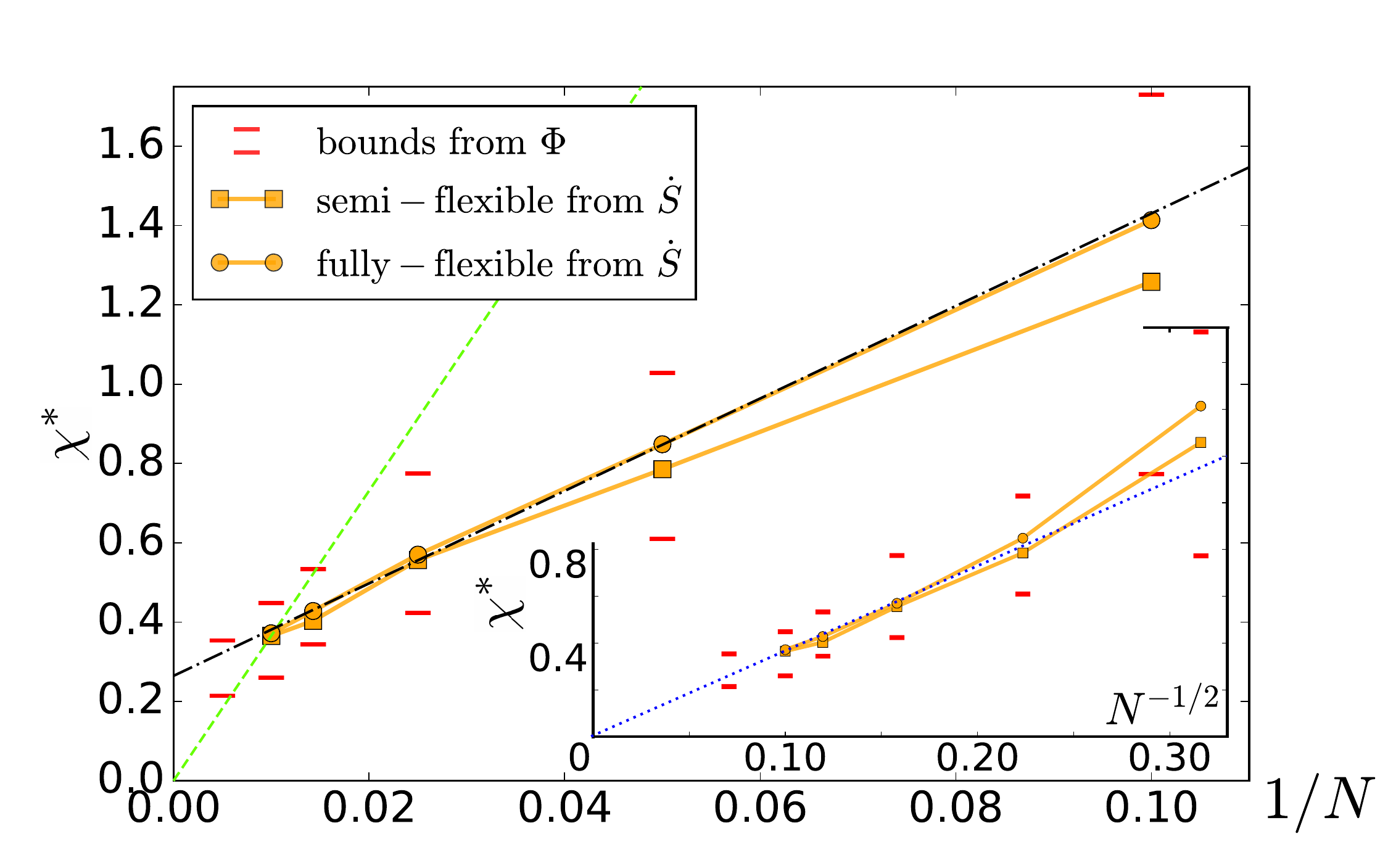}
\caption{(Colour online) Critical activity ratio as function of the polymer length. Critical activity ratio $\chi^{\ast}$ is plotted as function of $1/N$ for semi-flexible (squares) and fully-flexible (circles) chains. Red marks delimit the range of the largest gradient of order parameter $|\bar{\Phi}|$ extracted from Fig.~\ref{fig_phi_vs_T} (SM), symbols represent $\chi^{\ast}$ extracted from the entropy production rate $\dot{S}$ (see text). For $N=200$, the chains typically diffused only half of the box size, thus did not reach completely a steady state for all $\chi$. Based on the time profile of $|\Phi|$ (Fig.~S2 in SM) we marked the expected range of $\chi^{\ast}$. Green dashed line connects the origin with the point $\chi^{\ast}$ at $N=100$, black dot-dashed line is a linear function with an offset. Inset: the same data are plotted as function of $1/N^{1/2}$, blue dotted line connects the origin with the point $\chi^{\ast}$ at $N=100$.}
\label{fig_chi_vs_N}
\end{figure}

$(ii)$ Comparison of semi and fully-flexible systems shows a small decrease in the critical temperature ratio with the stiffness of the polymers (figures \ref{fig_phi_vs_T} and \ref{fig_chi_vs_N}). This is expected, as stiffer chains have a less pronounced correlation hole (chain self density), which leads to a slightly larger number of less favourable contacts \cite{chi_vs_stifness_Kumar}. 

$(iii)$ Simulations of systems with different $N$ (Fig.~\ref{fig_chi_vs_N}) exhibit the decrease of the temperature range of the transition with $N$. 

In simple terms, separation emerges because hot particles exhibit larger effective excluded volume and ``entrap'' passive particles as illustrated in \cite{Frey_2temp_phase_separation_PRL2016}. The hot particles exhibit higher pressure which is compensated by density change and phase separation \cite{Cates_Kardar_pressure_active_matter_NATURE,Joanny_Grosberg}. Still the mixing tendency is present and some chains diffuse in both temperature regions.

Commonly, one tracks phase separation by an order parameter based on the normalised number difference of the hot and the cold beads, $\Phi_{k} = (n_{h,k}-n_{c,k})/(n_{h,k}+n_{c,k})$ in the $k$-th sub-box (see SM for details). This is used in Fig.~\ref{fig_phi_vs_T}, where we plot $|\bar{\Phi}| (T_{h})$ i.e. $|\Phi_{k}|$ averaged over all sub-boxes and over steady state configurations and subtract $|\bar{\Phi}|(T_{h}=T_{c})$ that is not exactly zero due to finite size effects. 
Since we lack a classical ensemble we cannot employ the usual semi-grandcanonical scheme of exchanging chain types, allowing to evaluate the order parameter and higher order cumulants for the whole system \cite{Deutsch_Binder_Asymmetric_polymer_blends_phase_transition, Grest_Kremer_MD_asymmetric_melt_phase_transition, Binder_finite_size_scaling_MC_book}. This makes the $\Phi$ analysis suitable for trend description only. However, as we show below, the dependence of the critical temperature asymmetry on $N$ can be determined more accurately from the entropy production rate.

\begin{figure}[htb]
\includegraphics[width=\columnwidth]{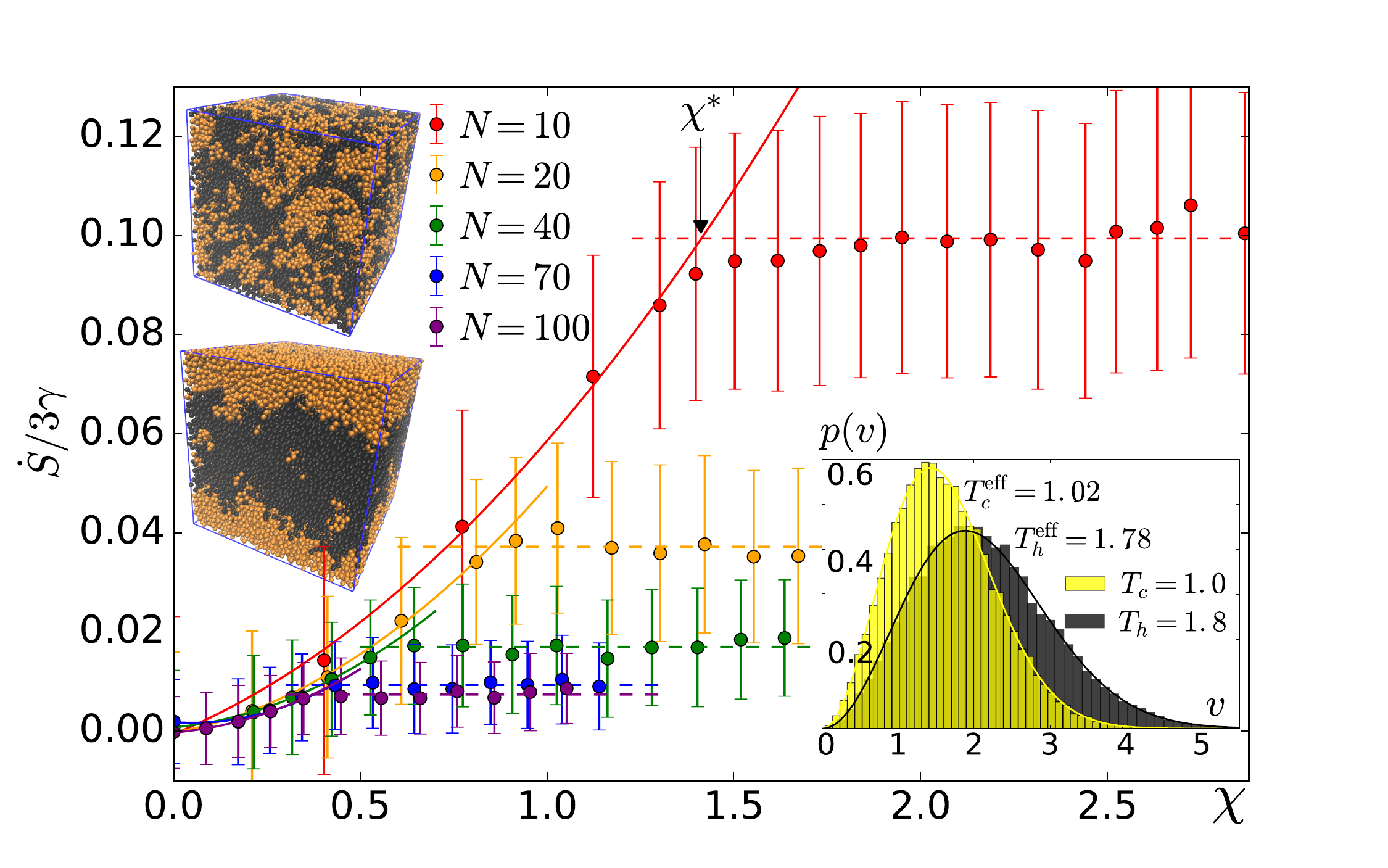}
\caption{(Colour online) Entropy production as an order parameter of the transition. Entropy production per hot particle in the steady state as function of $N$ and $\chi = (T^{\mathrm{eff}}_{h} - T^{\mathrm{eff}}_{c})/T^{\mathrm{eff}}_{c}$ for flexible polymers. The solid coloured lines are quadratic fit to the low $\chi$ (mixed) regime the dashed coloured lines are constant fit to the high $\chi$ (separated) regime. The black arrow show the extracted critical value of $\chi$ for $N=10$ as the crossover point between the two regimes. Error bars represent ensemble standard deviations as described in SM. Insets: Histograms - velocity distributions of the hot and cold particles in the separated state of $N=70$ and $T_{h}=1.8$, the effective temperatures are extracted as a fit of Maxwellian distribution. Two steady state $N=40$ system snapshots are shown: mixed (top) and phase separated (bottom) with hot chains (black) cold chains (yellow).}
\label{fig_entropy_prod}
\end{figure}

Particles thermalize much faster than the composition of their local environment changes. Thus both particle types have Maxwellian velocity distributions characterised by some effective temperatures $T^{\mathrm{eff}}_{h}$ and $T^{\mathrm{eff}}_{c}$ for hot and cold particles respectively, where $T_{h}>T_{h}^{\mathrm{eff}}>T_{c}^{\mathrm{eff}}>T_{c}$, as shown in the inset of Fig.~\ref{fig_entropy_prod}. This holds true for both, the mixed \emph{and} the phase separated systems (SM). The values of the effective temperatures, the energy flux between the hot and cold subsystems and the related entropy production rate are determined by the local particle environment, reflecting the order of the system. 

The Langevin thermostats act as reservoirs that pump energy to their respective subsystems by random force kicks and receive energy from them through the particle friction. The average power of the random force per hot particle is $\dot{E}^{\mathrm{rand}}_{h} = -3 \gamma T_{h}$ (we use units where $k_{B}=1$) and similarly for the cold reservoir, where the minus sign follows our convention that the flux is positive when flowing \emph{to} a reservoir. The friction force causes deceleration $\dot{\vec{v}} = -\gamma \vec{v}$ and therefore on average the related power per hot particle is $\dot{E}^{\mathrm{fric}}_{h} = \gamma m \langle v_{h}^{2}\rangle = 3 \gamma T^{\mathrm{eff}}_{h} $. The average is taken only over the $n_{i}$ particles connected to the thermostat $i$. In equilibrium $T^{\mathrm{eff}} = T$ and the energy flux due to random forces would be compensated by the friction. Out of equilibrium only the total power supplied to the system $- 3 \gamma (T_{c} n_{c} + T_{h} n_{h})$ has to be balanced in steady state by the amount dissipated by the friction forces i.e 
\begin{equation}
-T_{c} -T_{h} + T^{\mathrm{eff}}_{c} +T^{\mathrm{eff}}_{h} = 0
\label{power_balance}
\end{equation}
because the system is not changing or producing work. Here and in what follows we used that our systems have equal number of hot and cold particles ($n_{h}=n_{c}$).

In steady state the entropy production rate (per hot particle) of each reservoir amounts to the net heat flux to the reservoir per hot particle $3 \gamma (T^{\mathrm{eff}}_{i} -T_{i})$ divided by the temperature of the reservoir $T_{i}$ per unit time. The total entropy production of the system per hot particle $\dot{S}$ is the sum of the two contributions 
\begin{equation}
 \frac{\dot{S}}{3\gamma} = \frac{T^{\mathrm{eff}}_{h}}{T_{h}} + \frac{T^{\mathrm{eff}}_{c}}{T_{c}} - 2.
\end{equation}
The entropy production is always non-negative (equal to zero in equilibrium $T_{h} = T_{c}$) which follows from the temperature inequality and the power balance (\ref{power_balance}).

The analogue of the Flory interaction parameter is in our case the asymmetry of the \emph{effective} temperatures $\chi = (T^{\mathrm{eff}}_{h} - T^{\mathrm{eff}}_{c})/T^{\mathrm{eff}}_{c}$ in the steady state. The effective temperatures control the particles average kinetic and steric properties and therefore we choose $\chi$ over the thermostat temperature difference $dT=T_{h} - T_{c}$ (see also SM for analysis in terms of $dT$). Moreover, because of finite friction $\gamma$, a small $dT$ could lead to indistinguishable effective temperatures and the system would have no means to separate. Additionally, as the transition is governed by the effective temperature difference, the choice of friction should not have a strong effect on the transition position in terms of $\chi$, as opposed to $dT$.

The entropy production per hot particle in the steady state increases with $\chi$, due to the growth of the heat flux, until $\chi^{\ast}$, above which it plateaus as the interface between the phases develops (Fig.~\ref{fig_entropy_prod}). Therefore the critical activity ratio $\chi^{\ast}$ (symbols in figure \ref{fig_chi_vs_N}) is extracted as a crossover point between these two regimes. By phase separation, the hot chains maintain higher effective temperature by avoiding contacts with the cold chains, thus reducing the entropy production. 

As the entropy production per particle is an average quantity over all the particles it has the advantage over the simple estimate from the number difference as it is not affected by the finite size effects of the artificial division of the simulation box. On the other hand, in the separated regime, the entropy production in small systems is dominated by the interface and therefore the plateau value depends on the system size. In the infinite system size limit, the interface contribution vanishes and the entropy production per particle will be governed by the finite fraction of the hot chains interspersed in the cold phase (and vice versa). But even in thermodynamic limit the critical activity ratio would still be the crossover between two regimes (see SM). To quantify the interface effect, in SM we compare the entropy production per particle for two different system sizes and find that although the effect is relevant, the shift in the critical activity ratio is within our temperature precision. To overcome the difficulties, rather than going to larger system sizes it would be highly desirable to invent a method similar to semi-grandcanonical scheme for equilibrium phase transition. This we leave for future work.

In analogy to equilibrium phase separation, there is a competition between the tendency to mix arising from the heat baths proportional to the number of chains and 
the tendency to avoid unfavourable contacts proportional to the number of monomers, thus one could expect $\chi^{\ast} \sim 1/N$, i.e. $\chi^{\ast} N=const$. Interestingly, our data (Fig.~\ref{fig_chi_vs_N}) are not perfectly consistent with this trend. We approximately observe a linear ($1/N$) regime, but the extrapolation leads to a nonzero $\chi^{\ast}$ at infinite $N$. This would mean that for infinitely long polymers one needs a finite difference of effective temperatures to drive segregation, which seems counter-intuitive. In the inset of Fig.~\ref{fig_chi_vs_N} we plot the same data of $\chi^{\ast}$ as function of $N^{-1/2}$, which agrees with the trend for the longest chain lengths as well as the expected infinite chain length limit. There are several possible explanations for this observation:

$(i)$ Finite size effects and nucleation: Even for our relatively large systems the effect of the interface has a relevant contribution to the entropy production in the separated regime (see SM) and one can expect the offset to decrease in the infinite system size limit. Moreover, at constant $\chi N$ but very small $\chi$, nucleation processes might be necessary for the separation to occur \cite{Frey_2temp_phase_separation_PRL2016}. However, the velocity distributions overlap strongly at low $\chi$ and only a small fraction of hot particles is significantly hotter than the cold particles. Therefore their nucleation becomes unlikely, especially when the total number of particles is limited.

$(ii)$ Mixing tendency grows with $\chi$: The work \cite{Joanny_Grosberg} found for low densities that the separation of a system of spherical particles is governed by a free energy-like function with a term that favours mixing but grows with $\chi$. In other words, unlike in equilibrium phase transition, $\chi$ does not govern the separation tendency only. 

$(iii)$ Critical phase composition depends on $\chi$ (even for 50:50 mixture). In equilibrium phase transition with \emph{symmetric} interactions (i.e. interactions between like types of polymers are the same), the phase diagram is symmetric around the critical point located at the composition $1/2$ \cite{Rubinstein_book}. However, here the hot and cold phase have slightly different densities depending on $\chi$. At different $N$, we probe different temperature ranges (and partial densities) and therefore we look at a binodal at slightly different relative volume fractions. However, for equilibrium phase transition this effect should decrease the apparent $\chi_{c}$ (see SM). The asymmetry causes that our 50:50 mixture likely probes the critical point only in the limit of $N\rightarrow \infty$, which also raises metastability issues.
We briefly tested this for $N=40$ (see SM) and found the metastable region to be smaller than our temperature resolution. For longer chains however, this might be different.

The reduction of the entropy production rate in the separated state is interesting in the context of the minimum entropy production principle \cite{Prigogine_MEPP}. This, however, is applicable close to equilibrium only, while in our case the temperature differences and gradients take place on the particle scale and therefore can be quite large. Thus it would be interesting to look at much longer chains, and other models of activity especially Active-Ornstein-Uhlenbeck
Particles (AOUP) \cite{Cates_AOUP_PRL} where one controls the distance from equilibrium by persistence time characterising the velocity auto-correlation. For small persistence times one can define an effective temperature in a \emph{pure} AOUP system \cite{Maggi_Marini_colored_noise} and it exhibits vanishing entropy production \cite{Cates_AOUP_PRL}. Therefore it would be very interesting to see if the entropy production description also applies to passive-AOUP \emph{mixtures} or to what extent this would be violated by the non-Boltzmann distribution of the AOUPs. Another interesting and possibly relevant extension would be to incorporate velocity correlations along the chain \cite{Cugliandolo_active_dimers}. 

Consistent with symmetry considerations (SM), our data suggest that the entropy production scales as $\dot{S}\sim \gamma \chi^{2} M N$ for small $\chi$. Entropy produced on the relevant microscopic timescale $\gamma^{-1}$ i.e. $S\sim \chi^{2} M N$ competes with the mixing entropy proportional to $T M$ only. Here for small enough $\chi$ the constant $T$ should be close to average thermostat temperature and independent of $\chi$. If this applies, it results in $\chi^{\ast}\sim N^{-1/2}$ scaling consistent with our results.

Our computational study uncovers a novel effect in the activity-driven phase separation. The strong decrease of the critical activity with the length of the polymers makes the active process relevant in the chromatin (self)organization thus bridging the structure with the genome function. Moreover it facilitates the experimental exploration in artificial systems like \cite{Loewen_stat_mech_without_third_newtons_law} and could pave a way for new polymeric active materials. Already the present simplest model opens a fascinating branch in polymer physics, and allows one to use its tools to search for and understand new phenomena out of equilibrium.

\section{Acknowledgement}
We would like to thank Burkhard D\"{u}nweg, Kostas Daoulas, Tristan Bereau, Cristina Greco, Horacio Vargas, Torsten Stuehn and Alexander Grosberg for stimulating discussions.


%%%%%%%%%% Merge with supplemental materials %%%%%%%%%%
\pagebreak
\onecolumngrid
\begin{center}
\textbf{\large Supplementary material: Small activity differences drive phase separation in active-passive polymer mixtures}
\end{center}
\vskip 10pt
\twocolumngrid
%%%%%%%%%% Merge with supplemental materials %%%%%%%%%%
%%%%%%%%%% Prefix a "S" to all equations, figures, tables and reset the counter %%%%%%%%%%
\setcounter{equation}{0}
\setcounter{figure}{0}
\setcounter{table}{0}
\setcounter{page}{1}
\makeatletter
\renewcommand{\theequation}{S\arabic{equation}}
\renewcommand{\thefigure}{S\arabic{figure}}
\renewcommand{\bibnumfmt}[1]{[SR#1]}
\renewcommand{\citenumfont}[1]{SR#1}
%%%%%%%%%% Prefix a "S" to all equations, figures, tables and reset the counter %%%%%%%%%%

\section{Polymer model}
We use the standard bead-spring model \cite{KG_sup}: all beads interact with a purely repulsive shifted Lennard-Jones (LJ) potential $U_{\textrm{pair}}(r_{ij}) = 4 \epsilon [(\sigma/r_{ij})^{12}-(\sigma/r_{ij})^{6}]+\epsilon$ with a cutoff $r_{c}=2^{1/6} \sigma$. The chain connectivity is maintained by finitely extensible nonlinear elastic (FENE) potential $U_{\textrm{bond}}(r_{ij}) = -0.5 k R_{0}^{2} \ln [1 - (r_{ij}/R_{0})^{2}]$ for $r_{ij} < R_{0}$ between subsequent beads along the chain. 
In all simulations we use $k=30\epsilon/\sigma^{2}$, $R_{0}=1.5\sigma$. Additionally to the fully flexible chains, we study systems of semi-flexible chains with a bending potential $U_{\textrm{bend}} = k_{\theta} (1-\cos(\theta))$, where the $\theta$ is the angle between every two following bonds and $k_{\theta} = 1.5\epsilon$.

At first, we set up the chains as random walks with Lennard-Jones (LJ) and bond potential only between the nearest neighbors along the chains and then slowly turn on the LJ potential also for the other particle pairs to remove the overlaps. During the whole equilibration process we use a Langevin thermostat with $T=1.0 \epsilon$ (throughout we use units where $k_{B}=1$) and friction $\gamma = 1 \tau^{-1}$, where $\tau = \sigma \sqrt{m/\epsilon}$ is the natural time unit of the LJ interaction. Then the system was run about $10 N^{2}$ ($20 N^{2}$ for $N=100$), possibly with bending interaction, which was shown in \cite{Livia_equlibration_long_melts_sup} to be enough for a good equilibration for these lengths.

\section{Order parameter $\Phi$}
\begin{figure}[htb]
 \includegraphics[width=\columnwidth]{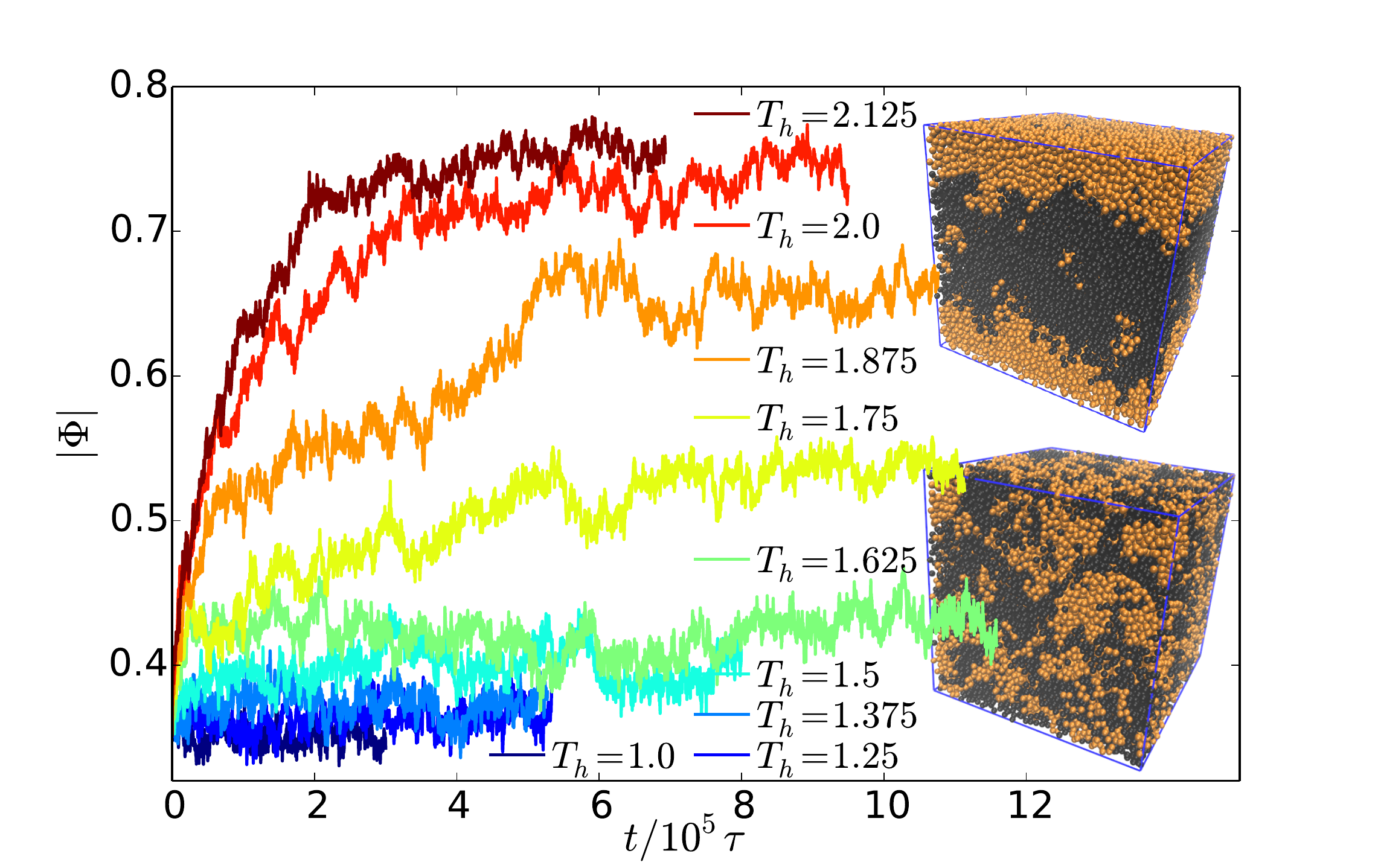}
 \caption{Order parameter as function of time for fully flexible melt with $N=40$ for different $T_{h}$. Insets: Snapshots of the phase separated (top) and mixed (bottom) system of the hot (black) and cold (yellow) chains.}
 \label{fig_order_parameter}
\end{figure}

The order parameter is based on the normalized number difference of the hot and cold beads, $\Phi_{i} = (n_{h,i}-n_{c,i})/(n_{h,i}+n_{c,i})$ in the $i$-th sub-box. In figure \ref{fig_order_parameter} we plot $|\Phi| = b^{-1} \sum_{i}^{b} |\Phi_{i}|$ i.e. $|\Phi_{i}|$ averaged over sub-boxes as function of time with typical early and late time system snapshots. Clearly, the sub-box size must be small enough to reduce the number of sub-boxes that probe the interfaces, but large enough to avoid big fluctuations due to small particle number effects. For our system sizes and temperature resolution $dT_{h} = 0.125 \epsilon$ we find similar results when $b=3^{3}$ to $b=10^{3}$ sub-boxes are used, however this method suffers from the finite size effects as mentioned in the main text. We circumvent this problem using the difference in the entropy production rate as an order parameter.

\begin{figure}[htb]
 \includegraphics[width=\columnwidth]{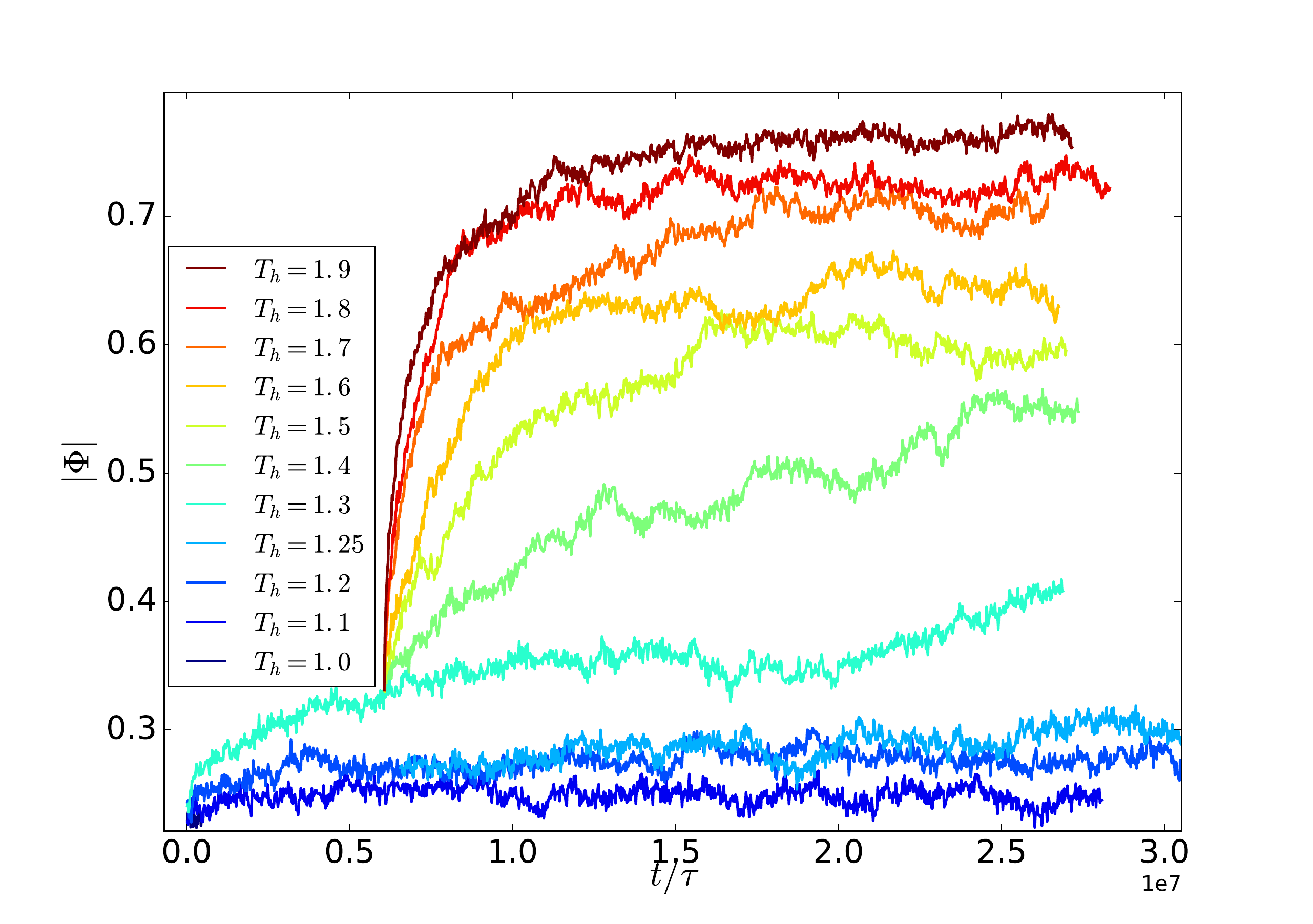}
 \caption{Order parameter as function of time for fully flexible melt with $N=200$ for different $T_{h}$. To save some computational time, already partially separating configurations with $T_{h}=1.3$ were used as the initial configurations for the higher temperatures. The time to reach the steady state and the observation time of the steady state are long (chains diffuse at least half the box size), however not all systems have reached the steady state (e.g.  $T_{h}=1.3$ due to its proximity to the critical point).}
 \label{fig_order_parameter_N200_vs_t}
\end{figure}

\section{Entropy production rate}
The steady state (when $|\Phi|$ as function of time plateaus) is typically reached after the chains diffused distance equal to the simulation box size.  
The normalized entropy production is calculated from the effective temperatures of the hot and the cold subsystem in the steady state. The effective temperatures are obtained as an average over at least forty conformations separated by $N^{2} \tau$ for chains of $N = 100$ (larger ensembles for shorter chains). The error bars in the entropy production (normalized by the number of hot particles) hence capture the ensemble variation due to conformational changes and related inherent velocity distribution of the particles of a given snapshot.

As mentioned in the main text the minimum entropy production principle (MEPP) \cite{Prigogine_MEPP_sup}, where the phase segregation steady state should be the one in which the entropy production is minimal under the external constraints is applicable to gentle deviations from the equilibrium only. In our systems the temperature differences and gradients can be quite large as they take place on the particle scale. Moreover close to equilibrium the entropy production is quadratic in the positions and velocities, while the mixed term is prohibited by requirement of non-negative entropy production upon time reversal (velocity changing sign). Also the entropy production must be positive if we interchange the roles of hot and cold reservoirs. As we showed, the position of the particle is coupled to the mean squared velocity of the local environment through the square of the effective temperature and therefore the validity of the MEPP is not guaranteed.

\section{Maxwellian velocity distribution in the mixed state}
\begin{figure}[htb]
\includegraphics[width=\columnwidth]{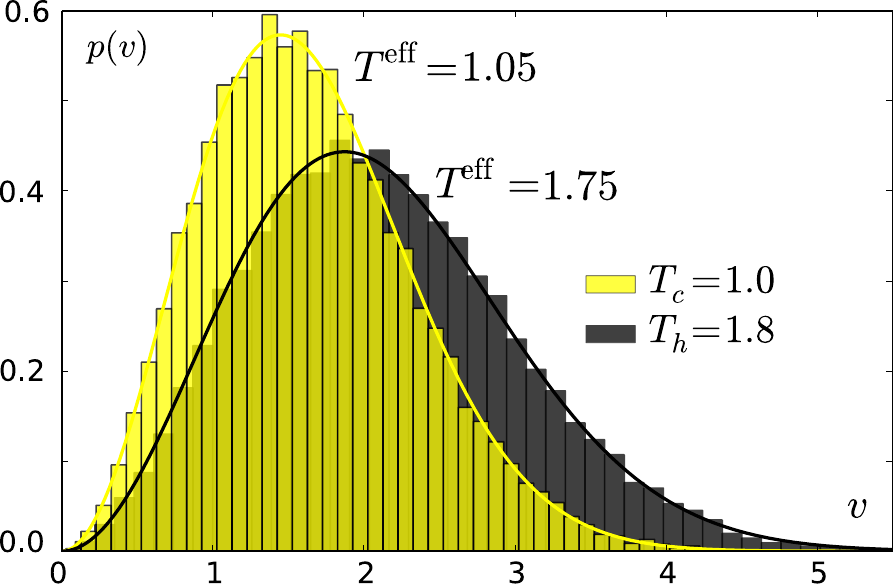}
  \caption{Velocity distributions of hot and cold subsystems in the early mixed state for $N=70$. Note the slight difference in the effective temperatures as compared to the effective temperatures in the separated state presented in the main text.}
\label{fig_vel_dist_sep}
\end{figure}

In separated state, it is expected to find a Maxwellian velocity distribution as most of the particles belong to a bulk of a phase with the same particle type. For completeness, in Fig.~\ref{fig_vel_dist_sep} we present a velocity distribution plot of the mixed state at time $\tau$ after the start of the simulation. This is long enough for the thermalization of the hot and cold subsystems (for our choice of $\gamma$). The velocity distributions are still Maxwellian, but the effective temperatures in the mixed state are further away from the thermostat temperatures as in the steady state.

\section{Effect of stiffness difference on separation}
Stiffness related effect arises from its dependence on temperature. In equilibrium systems, stiffness difference of polymer species favours the phase separation due to entropic effects \cite{chi_vs_stifness_polymer_blend_field_theory_Fredrickson, chi_vs_stifness_polymer_blend_simulation_Milner}. To check if this effect plays a role in our system, we ran independent simulations of equilibrium systems of a 50:50 mixture of fully and semi-flexible chains with $N=100$ at temperature $T_{c}=T_{h} = 1.0\epsilon$ and friction $\gamma = 1.0\tau^{-1}$. The ratio of the Kuhn segment lengths is in this case $b_{\textrm{semi}}/b_{\textrm{fully}} = 1.56$ \cite{Livia_equlibration_long_melts_sup}. This is much larger than the one arising from our temperature differences $\simeq 1.003$ (measured as the ratio of the mean bond lengths of hot and cold chains in fully flexible system with $T_{h} = 1.5$), however the equilibrium system remained well mixed. This is consistent with the findings of \cite{chi_vs_stifness_polymer_blend_simulation_Milner} for a slightly different polymer melt model. Therefore the flexibility difference of the hot and cold chains does not affect significantly the observed phase separation. 

\section{Critical phase composition dependent on $\chi$}
In symmetric equilibrium phase separation, the binodal has the form $\chi_{b} = \ln(\phi/(1-\phi))/N(2\phi-1)$ and the critical point is located at the composition $\phi =1/2$ \cite{Rubinstein_book_sup}. If for the different $N$ one finds the transition at a slightly different compositions $\phi = 1/2 + a/N$, where $a/N \ll 1/2$, one obtains a correction to the scaling of $\chi_{b}$ with $N$ as $\chi_{b} \sim A/N - |a|/N^{2}$. Extrapolating $\chi_{b}$ to infinite $N$ from small values of $N$ would lead to an negative offset, quite the opposite to what is observed in our data. 

\section{Metastability}
We briefly tested the metastability for $N=40$ chains by starting from a well separated conformation of $T_{h} = 2.5$, we switched the $T_{h}$ to values around the critical difference ($T_{h} = 1.375, 1.5, 1.625$ and $1.75$) and observed the evolutions. All the systems reached the same $|\bar{\Phi}|$ values as in the simulations that started from the mixed state. Therefore, the metastable region is likely to be smaller than our temperature resolution. For longer chains however, this might be different. To get the transition point more precisely, one would need to probe the different compositions, which we leave for future work.

\section{Entropy production as function of thermostat temperature differences}
\begin{figure}[htb]
\includegraphics[width=\columnwidth]{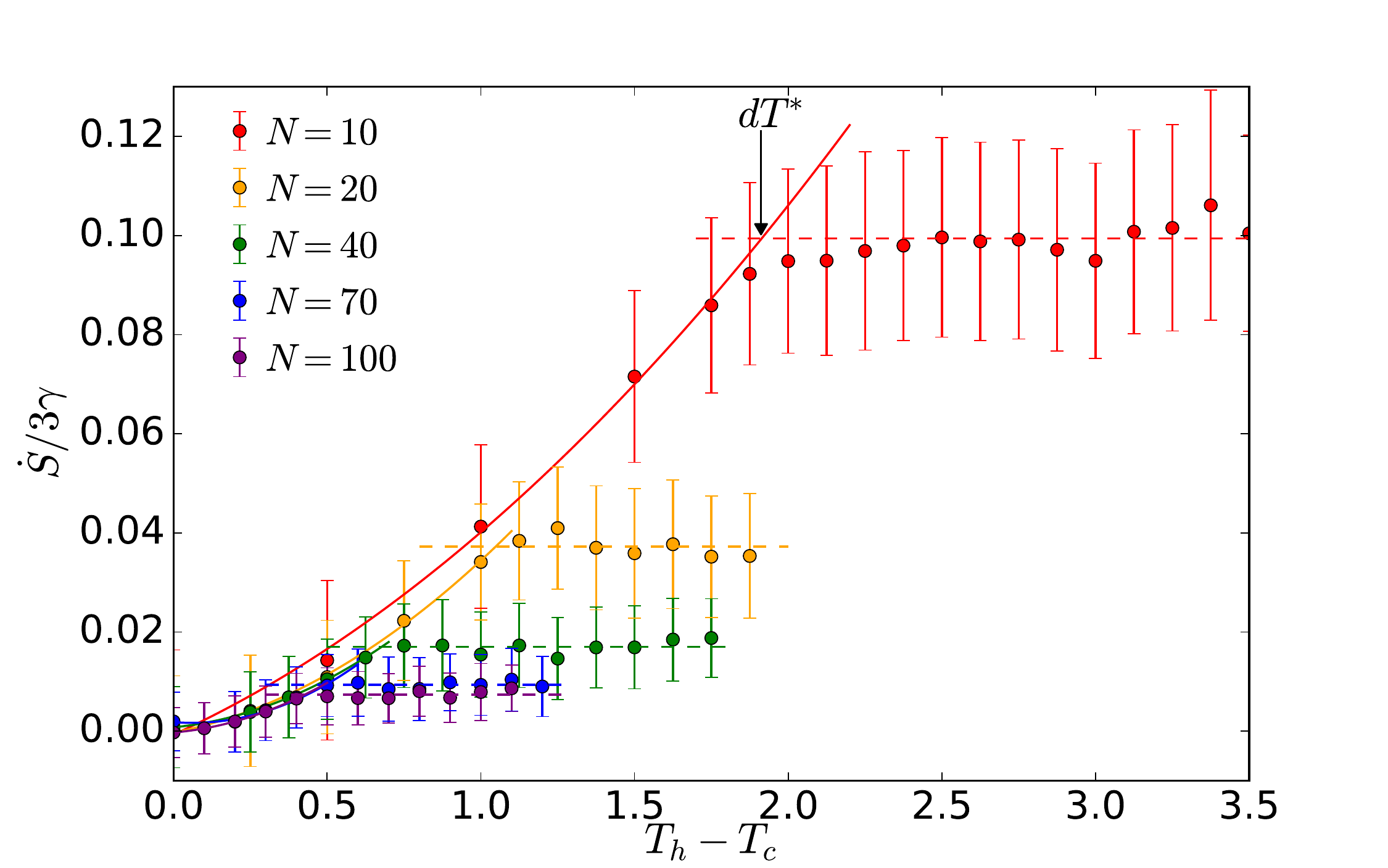}
   \caption{Entropy production as an order parameter of the transition. Entropy production per hot particle in the steady state as function of $N$ and $dT = T_{h} - T_{c}$ for flexible polymers. The solid coloured lines are quadratic fit to the low $dT$ (mixed) regime the dashed coloured lines are constant fit to the high $dT$ (separated) regime. The black arrow show the extracted critical value of $dT$ for $N=10$ as the crossover point between the two regimes. Error bars represent ensemble standard deviations.}
\label{fig_dotS_vs_dT}
\end{figure}

\begin{figure}[htb]
\includegraphics[width=\columnwidth]{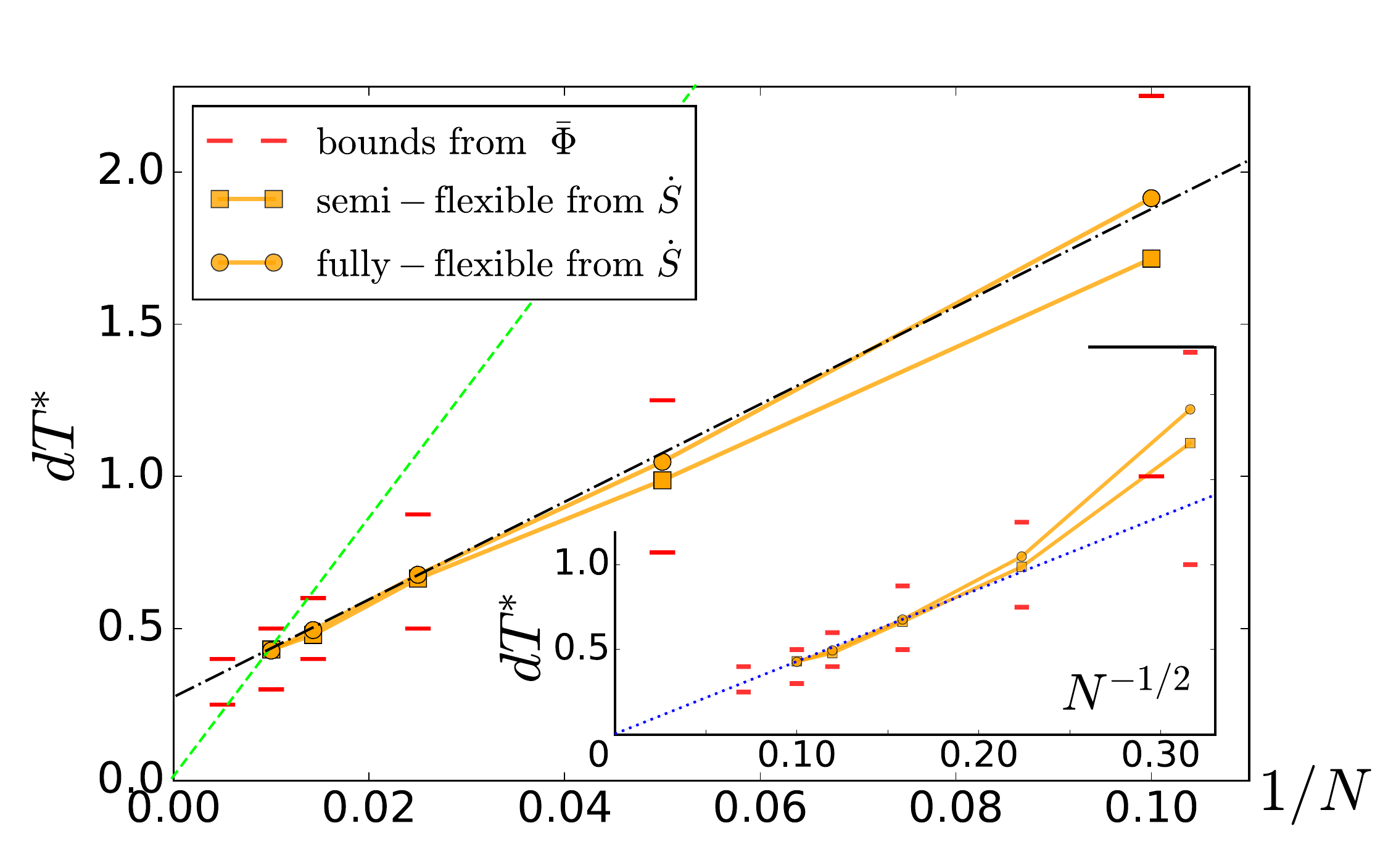}
  \caption{Critical thermostat temperature difference $dT^{\ast}$ is plotted as function of $1/N$ for semi-flexible (squares) and fully-flexible (circles) chains. Red marks delimit the range of the largest gradient of order parameter $|\bar{\Phi}|$ extracted from Fig. 1 (main text), symbols represent $dT^{\ast}$ extracted from the entropy production rate $\dot{S}$. Green dashed line connects the origin with the point $dT^{\ast}$ at $N=100$, black dot-dashed line is a linear function with an offset. Inset: the same data are plotted as function of $N^{-1/2}$, blue dotted line connects the origin with the point $dT^{\ast}$ at $N=100$.}
\label{fig_dTcrit_vs_N}
\end{figure}

As the temperature difference of thermostats $dT = T_{h} - T_{c}$ might be a natural experimentally interesting control parameter, in Fig.~\ref{fig_dotS_vs_dT} we plot the normalized entropy production as function of this quantity. The trend is the same as when plotted as function of $\chi$. Analogously as before, we extract critical temperature difference $dT^{\ast}$ and plot this as function of $1/N$ and $N^{-1/2}$ respectively in Fig.~\ref{fig_dTcrit_vs_N}. The offset is now more pronounced as the $dT$ is always greater than the effective temperature difference. As argued in the main text, the critical point $dT^{\ast}$ would depend more strongly on the choice of friction as in the case of $\chi^{\ast}$. 

\section{Finite size effects}
The finite size effects are of three sources: $(i)$ short chain lengths $N$, $(ii)$ entropy production in the separated regime is affected by the interface contribution and $(iii)$ fluctuations close to the $\chi^{\ast}$ are limited by the finite system size. 

To minimise the finite size effects, we used as large systems as possible while still being able to arrive at the steady states in reasonable computational time. As our results are consistent for $N\geq 20$, the short chain length effects are not significant for our main result that the critical activity ratio decreases with $N$. 

\begin{figure}[htb]
\includegraphics[width=\columnwidth]{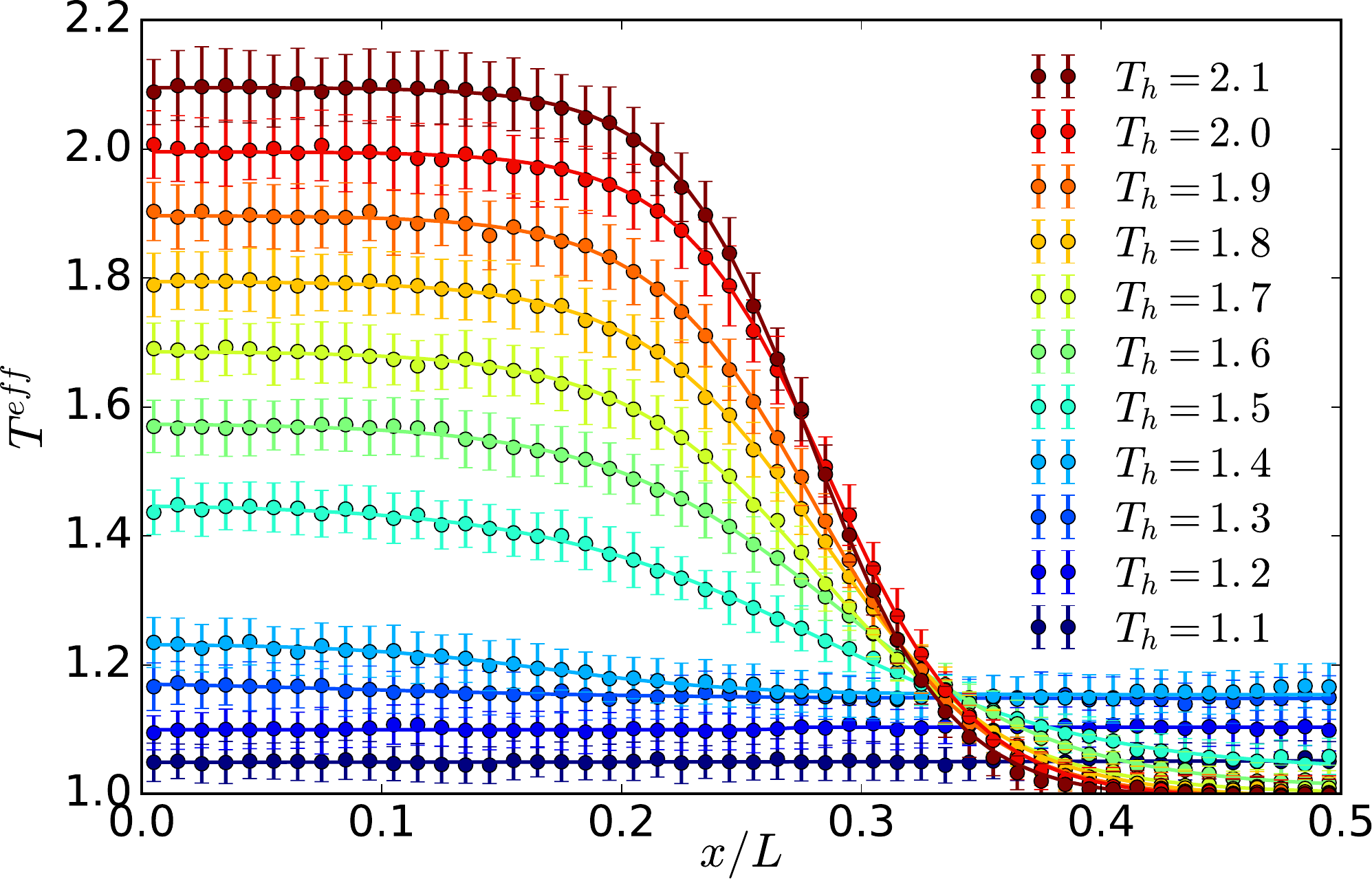}
  \caption{Effective temperature profile of $N=100$ fully-flexible system as function of a system coordinate $x/L$ that is perpendicular to the interface. The effective temperatures are an average over $100$ configurations from steady state separated by $N^{2}\tau$, the error bars represent ensemble standard deviation.}
\label{fig_temp_profile}
\end{figure}

The entropy production per particle in the phase separated regime has two contributions: $(a)$ contribution arising from the interface that spans the whole simulation box (of size $L$) and scales as $\dot{S}\sim L^{2}/(M N) \sim 1/(M N)^{1/3}$ and $(b)$ contribution arising from the hot chains interspersed in the bulk cold phase (and vice versa), which are independent of system size and scale with the number fraction of the two phases $\nu_{h}$ and $\nu_{c}$. In the thermodynamic limit, for finite $T_{h}-T_{c} = dT$ the first contribution vanishes, while the second remains finite. Then in the limit of $dT \to \infty$, the number fractions $\nu_{h}$ and $\nu_{c}$ go to zero. This can be seen in the figure \ref{fig_temp_profile} where effective temperature of a slab is plotted as function of the system coordinate perpendicular to the interface (if this exists). There for (high) values of $T_{h} \geq 1.7$ the effective temperature $T^{\mathrm{eff}}$ deep inside the hot phase  ($x/L \simeq 0$) is very close to the thermostat temperature. This shows that there are very few cold monomers in the hot phase. This suggest that even in thermodynamic and large $dT$ limit one can extract the the critical activity ratio as a crossover between two regimes - possibly a peak in the entropy production. A more detailed analysis of the interface width is left for a future work. 

\begin{figure}[htb]
\includegraphics[width=\columnwidth]{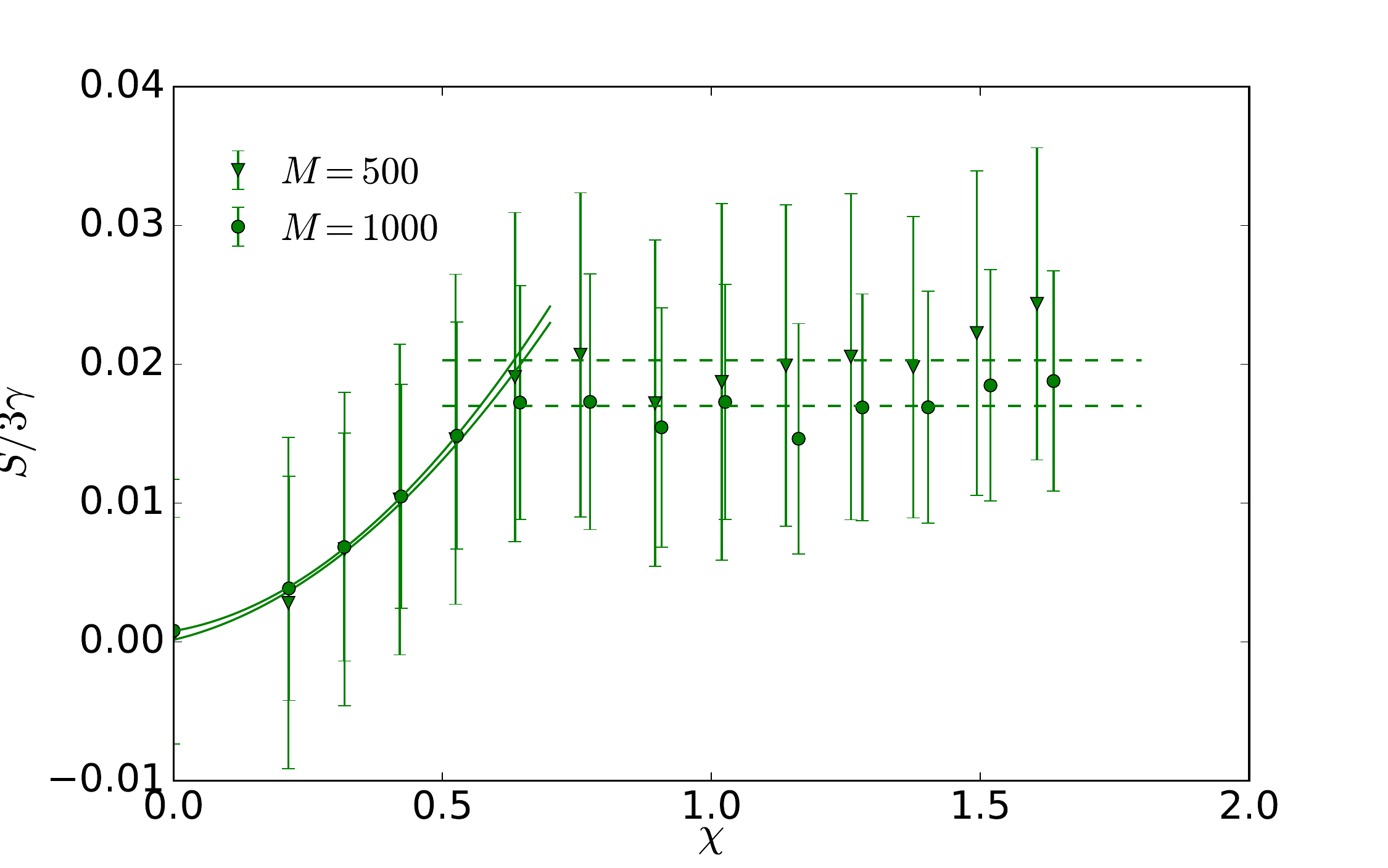}
  \caption{Comparison of entropy production per hot particle for two systems of $M=500$ and $M=1000$ chains with $N=40$.}
\label{fig_fse}
\end{figure}

To get an estimate of the extent of the finite size effects for our systems, we compared systems of $M=500$ chains and $M=1000$ chains for the intermediate chain length $N=40$. As shown in Fig.~\ref{fig_fse}, the mixed regime is very well reproducible as almost every monomer forms an interface between hot and cold constituents thus contributes to the entropy production. In the plateau regime, as expected the entropy production is lower for the larger system, which results in a shift of the critical activity ratio from about $\chi^{\ast} = 0.65$ to $\chi^{\ast} = 0.57$. This is a relevant shift, but comparable to our temperature resolution of $0.125\varepsilon$. 

The finite size effects related to critical fluctuations are a very well known problem in critical phenomena. As mentioned in the main text, in the case of equilibrium phase separation these can be partly circumvented by a semi-grandcanonical scheme, with finite size scaling of higher order cumulants of the order parameter. Analogous scheme for the present non-equilibrium situation is not yet known.

\end{document}